\documentclass[amsmath,amssymb,aps,10pt,prd,letterpaper,nofootinbib,balancelastpage,notitlepage,twocolumn,floatfix]{revtex4-2}
\usepackage{graphicx}	
\usepackage{ragged2e}	
\usepackage{bm}		    
\usepackage{slashed}
\usepackage{soul}
\usepackage[sort&compress]{natbib}	 	
	\setcitestyle{square,numbers,comma}	
\usepackage[colorlinks=true,urlcolor=blue,linkcolor=black,citecolor=red]{hyperref}	 
\usepackage{xcolor}
\usepackage{fontawesome5}

\usepackage{tikz-feynman}
\usepackage{braket}

\definecolor{orcidlogocol}{rgb}{0.65, 0.807, 0.223}
\newcommand{\orcid}[1]{\,\href{https://orcid.org/#1}{\textcolor{orcidlogocol}{\footnotesize\faOrcid}}}

\renewcommand{\eqref}[2][]{Eq{#1}.~(\ref{#2})}		

\begin{document}

\title{Cosmic Birefringence from CP-Violating Axion Interactions}
\date{\today}
\author{Xuheng Luo\orcid{0000-0002-5866-805X}}
\email{xluo26@jhu.edu}
\affiliation{The William H.~Miller III Department of Physics and Astronomy, The Johns Hopkins University, Baltimore, MD  21218, USA}
\author{Anubhav Mathur\orcid{0000-0003-0973-1793}\,}
\email{a.mathur@jhu.edu}
\affiliation{The William H.~Miller III Department of Physics and Astronomy, The Johns Hopkins University, Baltimore, MD  21218, USA}
\begin{abstract}
We explore the cosmic birefringence signal produced by an ultralight axion field with a small CP-violating coupling to bulk SM matter in addition to the usual CP-preserving photon coupling. The change in the vacuum expectation value of the field between recombination and today results in a frequency-independent rotation of the plane of CMB linear polarization across the entire sky. While many previous approaches rely on the axion rolling from a large initial expectation value, the couplings considered in this work robustly generate the birefringence signal regardless of initial conditions, by sourcing the field from the cosmological nucleon density. We place bounds on such monopole-dipole interactions using measurements of the birefringence angle from Planck and WMAP data, which improve upon existing constraints by up to three orders of magnitude. We also discuss UV completions of this model, and possible strategies to avoid fine-tuning the axion mass.
\end{abstract}

\maketitle


\section{Introduction}
\label{sec:intro}
Axions and axion-like particles (ALPs, hereafter also referred to as axions) comprise a well-motivated class of extensions to the standard model (SM) \cite{Peccei:1977hh,Weinberg:1977ma,Wilczek:1977pj,Frieman:1995pm,Caldwell:1997ii,Duffy:2009ig,Hui:2016ltb,Svrcek:2006yi,Arvanitaki:2009fg}. In addition to their role in solving the strong CP problem, they may serve as dark matter candidates or other new light (pseudo)scalars that generically couple to the SM \cite{Preskill:1982cy,Abbott:1982af,Dine:1982ah,Duffy:2009ig,Hui:2016ltb,Svrcek:2006yi,Arvanitaki:2009fg}.
The axion coupling to photons $-\frac{1}{4}g_{a\gamma}aF\tilde{F}$ has been widely studied in terrestrial, astrophysical, and cosmological contexts as a messenger of new physics \cite{Carroll:1989vb,Carroll:1998zi,Harari:1992ea,Carroll:1991zs,Chang:2023quo,Nguyen:2023czp,Lue:1998mq,Liu:2006uh,Ni:2007ar,Pospelov:2008gg,Finelli:2008jv,Galaverni:2009zz,Caldwell:2011pu,Gluscevic:2012me,Li:2013vga,Lee:2013mqa,Gubitosi:2014cua,Galaverni:2014gca,Gruppuso:2015xza,Planck:2016soo,Sigl:2018fba,Fedderke:2019ajk,Agrawal:2019lkr,Jain:2021shf,BICEPKeck:2021sbt,Jain:2022jrp,Eskilt:2022cff,Bortolami:2022whx,Galaverni:2023zhv,Liu:2019brz,Liu:2021zlt,DeRocco:2018jwe,Melissinos:2008vn,Obata:2018vvr,Liu:2018icu,Martynov:2019azm,Dekens:2022gha,Plakkot:2023pui,Gan:2023swl,Takahashi:2020tqv,Kitajima:2022jzz,Gonzalez:2022mcx}. One consequence of this interaction is that linearly-polarized photons travelling between two points with a varying axion field $a$ will experience a rotation in their polarization angle $\Delta \theta$ that is proportional to the difference in the field values $\Delta a$ at those points, independently of the photon path or frequency: $\Delta \theta = \frac{1}{2} g_{a \gamma} \Delta a$.

Cosmic birefringence refers to an apparent rotation of the plane of linear polarization of photons from the cosmic microwave background (CMB), seen across the entire sky. The effect can be quantified by decomposing the observed Stokes parameters $Q\pm iU$ in terms of $E$ and $B$ modes \cite{Komatsu_2022,Kamionkowski_1999}. The birefringence angle $\beta$ can be inferred from the power spectra of these modes, through the expression
\begin{align}
      \cos(4\alpha) C_\ell^{EB,\rm{obs}} &= \frac{1}{2}\sin{(4\beta)}(C_\ell^{EE,\rm{CMB}} - C_\ell^{BB,\rm{CMB}}) \nonumber \\
      &+ \frac{1}{2}\sin{(4\alpha)}(C_\ell^{EE,\rm{obs}} - C_\ell^{BB,\rm{obs}}) \nonumber \\
      &+ C_\ell^{EB,\rm{foreground}}
\end{align}
The intrinsic $EE$ and $BB$ auto-power spectra at the source, denoted CMB, can be predicted in $\Lambda$CDM. (We ignore the intrinsic $EB$ signal as it is too small to be detectable at present.) The miscalibration angle $\alpha$  accounts for uncertainties associated with the instrument as well as foregrounds (such as polarized dust emission) that may contribute to an observed $EB$ signal. Without independent knowledge of $\alpha$, observations of the CMB strongly constrain only the sum $\alpha + \beta$. This degeneracy can be broken by modelling the foreground polarization \cite{Minami:2019ruj}, which is only rotated by $\alpha$ as photons from the foreground travel a much shorter distance to the instrument compared to those from the CMB.

Non-zero isotropic cosmic birefringence is a generic expectation for an axion field coupled to photons with a vacuum expectation value (vev) which is homogeneous at cosmological distances and varies between recombination and today. As a result, observations of $\beta$ place constraints on any interaction that sources such a vev. In this work, we consider axions of mass $m_a$ coupled to both photons and SM nucleons, with the latter coupling $-g_{aN}a\bar{N}N$ responsible for sourcing a static monopole field of range $\sim 1/m_a$ around bulk matter $\left<\bar{N}N\right>$. This is known as the `monopole-dipole' scenario, and has been studied e.g. in the context of terrestrial searches for axion dark matter \cite{fedderke2023magpi,Agrawal:2023lmw}.
The axion coupling to nucleons violates CP symmetry. Small CP-violating (CPV) interactions are a reasonable expectation in new physics; they are already present in the SM \cite{ParticleDataGroup:2016lqr}, and are required in any explanation of baryogenesis \cite{Sakharov:1967dj,GAVELA_1994, Chang_2022}. For the case of a QCD axion, CPV interactions are expected to be induced from SM CPV interactions \cite{diluzio2021cpviolating,Pospelov:1997uv}, and can be independently probed in a variety of experiments \cite{PhysRevD.102.115026}. In our model, the CPV coupling allows for a large time-dependent axion vev (and hence an observable cosmic birefringence signal) to be sourced cosmologically even if the field has a small value at early times.

To our knowledge the most stringent measurement of cosmic birefringence arises from a joint analysis of data from Planck Public Release 4 \cite{Planck_intermediate} and the Wilkinson
Microwave Anisotropy Probe (WMAP) 9-year observations \cite{wmap}. An isotropic and frequency-independent signal of $\beta = 0.342^\circ{}^{+0.094^\circ}_{-0.091^\circ}$ (at the 68\% confidence level) has been reported \cite{Komatsu_2022} (here, a positive value refers to a clockwise rotation). This has prompted possible explanations such as Faraday rotation of CMB photons due to primordial magnetic fields \cite{Polarbear_2015}, scalar fields acting as early dark energy that also couple to electromagnetism \cite{Komatsu_2023}, and others \cite{Khodagholizadeh:2023aft,Nakai:2023zdr,Geng:2007va,Zhou:2023aqz,Nakagawa:2021nme}. 

In this work, we determine the limits on axion monopole-dipole couplings from measurements of cosmic birefringence. The axion is not required to have any initial abundance as dark matter or (early) dark energy \cite{Gasparotto_2022}. Its mass cannot be substantially greater than the scale of the horizon at recombination ($m_a \lesssim 10^{-27} \text{ eV}$) to ensure that the field is homogeneous and does not oscillate at early times. We also map out the parameter space explaining the apparent observation of non-zero $\beta$, which has a current significance of $3.6\sigma$. The required couplings are weaker than current bounds, and may be accessed by future axion experiments \cite{Graham:2015ouw} and tests of the equivalence principle (EP) \cite{Brzeminski:2022sde,Graham:2015ifn}.

The remainder of this paper is organized as follows. 
We describe our effective model in Sec~\ref{sec:Model}, and compute the resulting axion evolution and birefringence signal in Sec.~\ref{sec:results}. We subsequently discuss the early-time evolution of the model and possible UV completions in Sec.~\ref{sec:UV}, and conclude in Sec.~\ref{sec:Conclusion}.

\section{Model}\label{sec:Model}
We consider the following Lagrangian describing axions $a$ with a coupling to photons alongside CP-violating (CPV) interactions:
\begin{equation}\label{lagrangian1}
    {\cal L}\supset -\frac{1}{2}D_\mu a D^\mu a +\frac{1}{2}m_a^2a^2 -\frac{1}{4}g_{a\gamma}aF\tilde{F} + \mathcal{L}_{\rm{CPV}},
\end{equation}
where $D_\mu$ is the covariant derivative in an expanding Universe. Without specifying the UV physics, many CPV interaction terms can appear in the IR\footnote{Even if such interactions are absent in the bare Lagrangian, they are expected to be generated from SM CPV interactions at some level.}. We take the following as a benchmark:
\begin{equation}\label{lagrangian2}
    {\cal L _{\rm{CPV}}} = -g_{aN}a\bar{N}N
\end{equation}
In the low mass regime, this operator gives rise to a fifth force between nucleons which is strongly constrained by EP tests. The current best constraints, from the MICROSCOPE satellite, are $g_{aN} < 9\times 10^{-25}$ \cite{Microscope_2022}. At these parameter values, the fifth force is much weaker than gravity and so leaves the structure of nucleons in our Universe unaffected.

Through $-g_{aN}a\bar{N}N$, the nucleons in the Universe source an axion vev that is proportional to the baryon number density, since $\braket{\Bar{N}N} \approx \braket{\Bar{N}\gamma^0 N} = n_N$. In terms of cosmological parameters,
\begin{equation}
\braket{\Bar{N}N} \approx n_{N,0} (1+z)^3 = \frac{3\Omega_b H_0^2 m_{\rm{pl}}^2}{8\pi m_N}(1+z)^{3} 
\end{equation}
where $n_{N,0}$ is the nucleon number density today, $\Omega_b$ is the cosmological fraction of baryon energy density, $H_0$ is the Hubble constant today and $m_N \sim 1\text{ GeV}$ is the average nucleon mass.

\section{Results}
\label{sec:results}

In a cosmologically evolving axion background, CMB photons experience a rotation in their linear polarization angle between emission and absorption due to the presence of the axion photon coupling in Eq.~\ref{lagrangian1}. As worked out e.g. in Ref.~\cite{Fedderke_2019}, the net rotation $\Delta \theta$ is independent of the path of the photons and depends solely on the difference in field values $\Delta a$:
\begin{equation}
    \Delta \theta = \frac{1}{2} g_{a\gamma} \Delta a = \frac{1}{2} g_{a\gamma} (a(0) - a(z_{\rm rec}))
\end{equation}
where $z_{\rm rec}\approx 1080$ is the redshift at recombination. It is precisely this rotation that is measured as the birefringence angle $\beta$, which we compare to the predicted $\Delta \theta$ in absolute value since our effect can easily be reversed by changing the sign of the photon coupling. At the present limit for $g_{a\gamma}$, the requirement to achieve $\beta \sim 0.1^\circ$ is $|\Delta a| \sim 10^9\text{ GeV}$.

In our analysis we treat the axion field as being homogeneous and of solely cosmological origin. While an axion vev can also be sourced astrophysically through the same coupling, this is always negligible in the parameter space we consider\footnote{At the position of the Earth today, the dominant local source of axions is the Milky Way, responsible for a vev of size $a_{\text{MW}}\sim 10^5\text{ GeV} (g_{aN}/10^{-27})$.}. Moreover, $\delta \rho / \rho \sim 10^{-5}$ at recombination, meaning that fluctuations in the axion vev across the sky (which would in principle induce spatial correlations in the photon polarization) are unobservably small.

The axion field satisfies the following equation of motion:
\begin{equation}\label{axionbg}
       \ddot{a} + 3H\dot{a} + m_a^2 a= -g_{aN}\braket{\Bar{N}N},
\end{equation}
where dots denote the derivative w.r.t. the proper time.
When solving for the dynamics of $a$, the initial value $a_i$ is set to zero after Big Bang nucleosynthesis (BBN). (The initial $\dot{a}$ is unimportant due to the large Hubble friction.) The subsequent evolution has a negligible dependence on the choice of $a_i$, provided that it is smaller than the peak value $a_\text{max}$ attained in the $a_i = 0$ case, which is $\gtrsim 10^9\text{ GeV}$ in the parameter regime of interest. (The plausibility of this early-time behavior and the possible UV completions is discussed in Sec.~\ref{sec:UV}.) Conversely, when $a_i\gtrsim a_{\text{max}}$, the source term does not contribute appreciably to the cosmological evolution of $a$. Ref.~\cite{Komatsu_2023} treats the cosmic birefringence signal in this initial-condition dominated scenario. The presence of the CPV interaction term ensures that in the mass range $m_a \lesssim 10^{-27}\text{ eV}$, there is a substantial birefringence signal regardless of the size of $a_i$, which is a unique feature of this work.

\begin{figure}
    \centering
    \includegraphics[width=0.95\columnwidth]{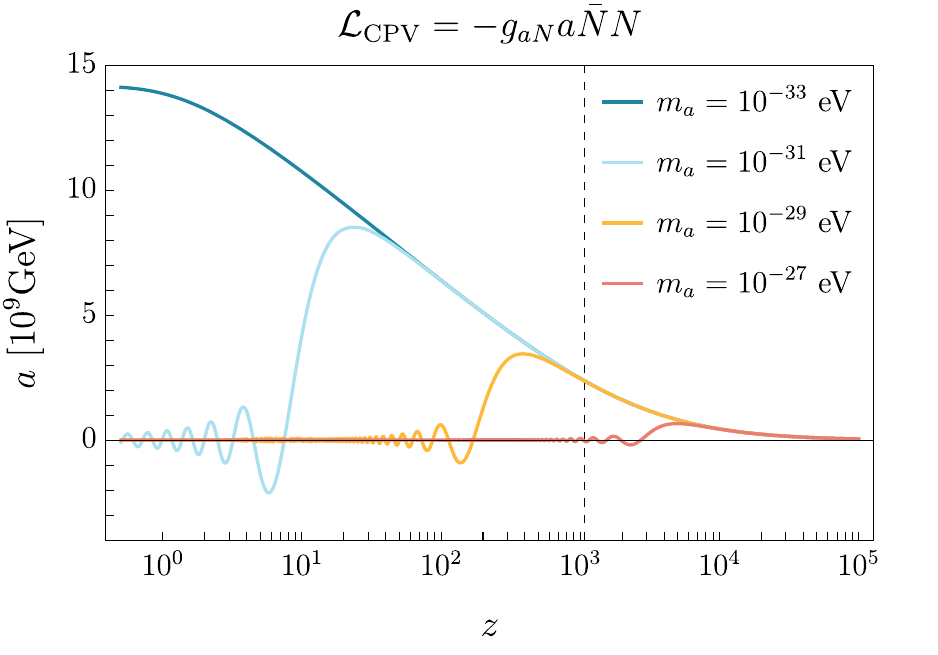}
    \caption{The evolution of the axion field $a$ for different choices of the mass $m_a$, determined numerically by solving Eq.~\ref{axionbg}  with initial condition $a_i = 0$. The field is sourced by a CP violating coupling to nucleons (Eq.~\ref{lagrangian2}), with $g_{aN} = 10^{-27}$. The dashed line corresponds to the redshift of recombination.}
    \label{fig:field-evol}
\end{figure}

Analytically, the behaviour of the axion field for the $a_i = 0$ case may be estimated as follows. While $m_a \lesssim 3H(z)$, the axion vev grows alongside the number of constituents within the horizon $\sim1/H(z)$. The growth is $\propto 1/z$  during radiation domination and subsequently slows, evolving for $z\lesssim z_{\rm{MRE}}$ as
\begin{equation}
    a_{\rm{grow}}(z) \approx \frac{g_{aN} n_{N,0}}{H_0^2} \left[\frac{8(\frac{1+z}{z_{\rm{MRE}}})^{3/2} - 12\log(\frac{1+z}{z_{\rm{MRE}}}) + 1}{18\Omega_m}\right]
\end{equation}
where $z_{\rm{MRE}}\approx\Omega_m/\Omega_r$ is the redshift of matter-radiation equality.
At the critical redshift $z_*$ defined by $m_a = 3H(z_*)$, the field value reaches a maximum of $a_{\rm max} \approx a_{\rm{grow}}(z_*)$ and begins to oscillate. Due to Hubble friction, the size of the peaks decreases with each oscillation, with the upper envelope following
\begin{equation}\label{eq:envelope}
    a_{\rm{env}}(z) \approx a_{\rm{max}} \left( \frac{z}{z_*} \right)^{3/2} 
\end{equation}
This means that for a fixed coupling strength, the largest $|\Delta a|$ (and hence the strongest birefringence signal) is achieved if the axion does not begin to oscillate much earlier than recombination i.e. $m_a \lesssim 3H(z_{\rm rec})$.
The numerical evolution of $a$ is shown for a range of masses in Fig.~\ref{fig:field-evol}. Notice that the field value at recombination is independent of mass as long as $m_a \lesssim 3H(z_{\rm rec})\sim 10^{-28} \text{ eV}$, and that $a(0) \ll a(z_{\text{rec}})$ for $m_a \gtrsim 3H_0 \sim 4\times 10^{-33}\text{ eV}$.

\begin{figure}
    \centering
    \includegraphics[width=0.95\columnwidth]{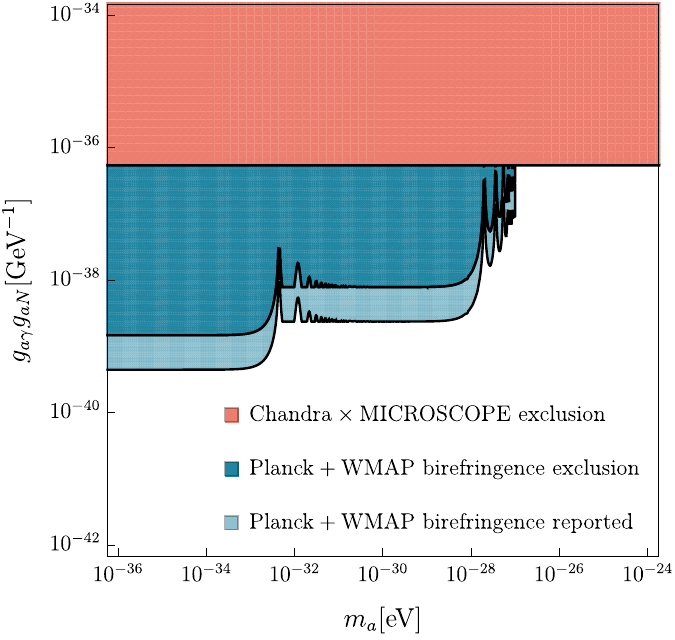}
    \caption{Parameter space for the product coupling $g_{a\gamma}g_{aN}$ governing the strength of axion monopole-dipole interactions. Existing bounds, determined by combining the X-ray spectral distortion constraints on $g_{a\gamma}$ from Chandra \cite{Reynolds:2019uqt,Reynes:2021bpe} and the fifth-force constraints on $g_{aN}$ from MICROSCOPE \cite{Microscope_2022,PhysRevD.102.115026}, are shown in red. The dark blue region is ruled out by measurements of the birefringence angle in Ref.~\cite{Komatsu_2022}, and the light blue region corresponds to the parameter values sufficient to explain the non-zero birefringence angle reported by the same analysis (both at the 95\% confidence level).
 }
    \label{fig:par-space}
\end{figure}

The parameter space probed by observations of cosmic birefringence is shown in Fig.~\ref{fig:par-space}. This result improves on existing constraints on the product coupling $g_{a\gamma}g_{aN}$ by up to three orders of magnitude. There are three distinct regions in the plot. For $m_a \lesssim 3H_0$, the axion is light enough that it never starts oscillating---in this case the field continuously grows, from $a_{\rm{rec}} \approx a_{\rm{grow}}(z_{\rm{rec}})$ at recombination to $a_{\rm{tod}} \approx a_{\rm{grow}}(0)$ today, resulting in a predicted birefringence angle of size
\begin{align}
    |\Delta \theta| &\approx \frac{1}{2}g_{a\gamma}(a_{\rm{tod}}-a_{\rm{rec}}) \qquad \qquad \, [m_a \lesssim 10^{-33} \text{ eV}] \\ 
    &\approx 0.4^{\circ}\left(\frac{g_{a\gamma}}{10^{-12}\text{ GeV}^{-1}}\right)\left(\frac{g_{aN}}{10^{-27}}\right)
\end{align}
For $3H_0 \lesssim m_a \lesssim 3H(z_{\rm{rec}})$, the axion attains a fixed value $a_{\rm{rec}}$ at recombination and subsequently begins oscillating, decreasing to a negligible value by today. The corresponding birefringence signal is 
\begin{align}
    |\Delta \theta| &\approx \frac{1}{2}g_{a\gamma}a_{\rm{rec}} \qquad \:[10^{-32}\text{ eV} \lesssim m_a \lesssim 10^{-28}\text{ eV}] \\
    &\approx 0.08^{\circ}\left(\frac{g_{a\gamma}}{10^{-12}\text{ GeV}^{-1}}\right)\left(\frac{g_{aN}}{10^{-27}}\right)
\end{align}
Notice that in each of the above regimes, the result is independent of the axion mass.

Finally, for $m_a \gtrsim 3H(z_{\rm{rec}})$, the axion field peaks and begins oscillating before recombination, leading to a degradation of the signal which can be estimated using Eq.~\ref{eq:envelope}:
\begin{align}
    |\Delta \theta| &\approx \frac{1}{2}g_{a\gamma}a_{\rm{env}}(z_{\rm{rec}}) \; \;[10^{-28}\text{ eV} \lesssim m_a \lesssim 10^{-27}\text{ eV}] \mkern-100mu \\ 
    &\approx 0.02^\circ \left(\frac{g_{a\gamma}}{10^{-12}\text{ GeV}^{-1}}\right)\left(\frac{g_{aN}}{10^{-27}}\right) \left(\frac{m_{a}}{10^{-28}\text{ eV}}\right)^{-1.06}
\end{align}
We choose to cut off the region above $10^{-27}\text{ eV}$, where numerous oscillations of the axion field during the CMB decoupling epoch leads to a ``washout" of the field value as described in Ref.~\cite{Fedderke_2019}\footnote{The reduction of the net polarized fraction of the CMB due to this effect places limits which are weaker than existing bounds, so they are not included here.}.

\section{Discussion}\label{sec:UV}

While the birefringence signal described in the previous section is robust to the choice of initial conditions, the late-time evolution of the axion field is dominated by the CP-violating term only when its value $a_i$ after BBN is small compared to the field excursion $\sim 10^9\text{ GeV}$ relevant to our parameter space. This motivated the choice of $a_i = 0$ in the earlier analysis. It is worth considering whether a small initial value is generically expected from the early-time evolution of the model presented in Eq.~\ref{lagrangian1}.

Analogous to misalignment in the case of the QCD axion, a compact potential with period $f_a$ will prevent $a_i$ from dominating the evolution provided that $f_a \sim a_{\rm{max}}$.
An even smaller early-time axion vev $a \ll f_a$ appears in some models of the early Universe \cite{Di_Luzio_2020}. One possibility is for the axion to have a mass during a period of low-scale inflation \cite{Graham:2018jyp,Takahashi:2018tdu}.
As long as inflation lasts sufficiently long, the scalar will slow roll towards the minimum of its potential, resulting in a suppressed $a$.
Other models predict a small $a$ even when the axion potential appears after inflation \cite{Baratella_2019,Nakagawa:2021nme}.

Due to the presence of the nucleon coupling, there is additional model dependence in the evolution of $a$ after inflation, as the field may grow to a large $a_i$ by the end of BBN. Above the critical temperature of the QCD phase transition $\Lambda_{\rm{QCD}}$, the nucleons are no longer the relevant degree of freedom and the effective interaction in Eq.~\ref{lagrangian2} breaks down. This complication can be avoided by requiring that the universe reheat to a temperature below $\Lambda_{\rm{QCD}}$, in which case the number density of nucleons is Boltzmann suppressed throughout, and a small $a_i$ is maintained. Further possible interactions arise at the loop level \cite{Kapusta:2006pm,Brzeminski:2020uhm,Gan:2023wnp}, notably to electromagnetism through $\sim (\alpha_{\rm{EM}}g_{aN}/6\pi m_N)  aFF$ (in the absence of additional UV contributions to this operator). Even in the presence of a large number density of charged leptons at early times, we have verified that the thermal contribution to the axion vev is negligible.\footnote{A CP-violating coupling to photons can also modify fundamental parameters such as $\alpha_{\rm{EM}}$ in the presence of a large expectation value for the axion field. This is well below the threshold of detection \cite{Kaplan:2022lmz,Stadnik:2015kia,Hart:2019dxi,Sibiryakov:2020eir,Bouley:2022eer}.}

Having addressed the evolution of the axion field prior to BBN, we now specify a renormalizable theory that generates the effective interactions of Sec.~\ref{sec:Model}. The UV Lagrangian is
\begin{equation}\label{UV_lagrangian}
    {\cal L}_{\rm UV}\supset \epsilon e A_\mu\Bar{\psi}\gamma^\mu\psi  - y_{\rm{even}}a\Bar{\psi}i\gamma_5 \psi - y_{\rm{odd}}a\Bar{\psi} \psi.
\end{equation}
Here, $\psi$ is a heavy fermion with mass $m_\psi \gtrsim 10\text{ TeV}$ to avoid collider constraints, carrying a charge $\epsilon e$ under SM electromagnetism. $\psi$ also couples to a light scalar $a$, with both CP-conserving and CP-violating interactions. Upon integrating out the heavy fermion, the following interactions are produced:
\begin{equation}\label{eq:L_IR}
    {\cal L _{\rm{IR}}} \sim  \frac{1}{4}\frac{y_{\rm{even}}\epsilon^2e^2}{4\pi^2m_\psi} aF\Tilde{F} + \frac{1}{4}\frac{y_{\rm{odd}}\epsilon^2e^2}{6\pi^2m_\psi} aFF.
\end{equation}
The latter term further generates axion-nucleon interactions $- g'_{aN} a\Bar{N}N$ from the nucleon binding energy \cite{Damour_2010} as well as from loop interactions \cite{Dupays_2007}. This results in an effective nucleon coupling of strength \cite{Microscope_2022}
\begin{equation}\label{dilaton}
g'_{aN}  = -Q_e m_N \times  \frac{y_{\rm{odd}}\epsilon^2e^2}{6\pi^2m_\psi},
\end{equation}
where $Q_e$ is the dilaton charge of the nucleon, which for Hydrogen is $\approx 6.8 \times 10^{-4}$ \cite{Damour_2010}. 
The nucleon coupling
of interest requires a large CPV photon coupling which can spoil the small initial condition $a_i$ depending on the early universe history. Here, we only discuss its feasibility as a UV completion.

An ultralight scalar field generically suffers from large loop corrections to its mass unless additional symmetry is imposed to protect it \cite{Dimopoulos:1981zb,Randall:1999ee,Arkani-Hamed:1998jmv,Graham:2015cka,Susskind:1978ms,Weinberg:1975gm}. In axion models, this is accomplished by the shift symmetry of the axion field. It can be shown that through a field-dependent redefinition of $\psi$, the CP-conserving term in Eq.~\ref{UV_lagrangian} may be rewritten as a derivative coupling in $a$ \cite{Quevillon:2021sfz}. As a result, corrections to the axion mass in Eq.~\ref{UV_lagrangian} arise exclusively from the CP-violating sector, which breaks the shift symmetry explicitly in both the UV and effective theories.
In the absence of fine-tuning, the axion mass is corrected to 
\begin{equation}
    \Delta m_a^2 \sim y_{\text{odd}}^2 \Lambda^2.
\end{equation}
where $\Lambda$ is the cut-off scale of this theory. Setting $\Lambda \sim m_\psi \gtrsim 10 \text{ TeV}$, the mass is very unnatural in this model.
However, the severity of this problem depends heavily on the UV theory: as in the case of the QCD axion, CPV interactions may originate from a distinct sector such that they do not significantly correct the scalar mass in Eq.~\ref{UV_lagrangian}.
Another possible resolution is to impose a discrete symmetry which relaxes the fine-tuning of the mass \cite{Hook_2018}. Motivated by this idea, a number of natural particle models have been built for e.g. a scalar with a Yukawa interaction \cite{Hook_2018,Brzeminski:2020uhm,Gan:2023wnp} and an ultralight QCD axion \cite{Luzio2021:2102.00012v2}. We schematically show how to do so for $\mathcal{L}_{\rm UV}$ by enhancing it to a $\mathbb{Z}_N$-protected theory.

To realize this symmetry, we require $N$ copies of the SM and $N$ fermions $\Psi_k$ with identical mass $M_\Psi$ and charge $\epsilon e$. We additionally introduce a complex scalar $\Phi$ that is a SM singlet, with a Mexican hat potential $V(|\Phi|)$. The axion will emerge as the Goldstone boson associated with this field.
The fermions are initially massless, and interact with the scalar through
\begin{align}\label{Zn_lagrangian}
    {\cal L}_{N} \supset & \sum_{k = 0}^{N-1} (Y_{\text{even}}e^{2\pi i k/N}\Phi\bar{\Psi}_k P_R \Psi_k + \text{h.c.}) \\
     + &\sum_{k = 0}^{N-1}  (iY_{\text{odd}}e^{2\pi i k/N}\Phi\bar{\Psi}_k \Psi_k + \text{h.c.}) \nonumber
\end{align}
which is symmetric under the transformations
\begin{align}
    \mathbb{Z}_N: \Psi_k &\rightarrow \Psi_{k+1 (\rm{mod} \ N)}\\
    \Phi &\rightarrow e^{2\pi i /N}\Phi.
\end{align}
In the limit $Y_{\text{odd}}\rightarrow 0$, the theory also exhibits axial $U(1)$ symmetry 
which is spontaneously broken due to $V(|\Phi|)$. Upon acquiring a vev $f_a$, the scalar can be rewritten as $\Phi = f_a\exp(ia/f_a)/\sqrt{2}$.
Expanding $\mathcal{L}_N$ to linear order in $a$, $\mathcal{L}_{\rm UV}$ is recovered as the $k=0$  term, which we take to be our copy of the world. The fermions are now heavy, with $M_\Psi = Y_{\text{even}}f_a/\sqrt{2}$, and can be integrated out. This results in the desired effective theory (Eq.~\ref{eq:L_IR}), with quantum corrections to the axion mass expected to be relaxed exponentially by $\sim(Y_{\rm{odd}}f_a/M_\Psi)^{N-2}$ \cite{Brzeminski:2020uhm,Gan:2023wnp}, though the detailed computation is deferred to future work. A natural scalar mass is attained throughout the parameter space of interest for $N \gtrsim 7$.

\section{Conclusion}\label{sec:Conclusion}

Cosmic birefringence is widely appreciated as a simple observational signature arising from the coupling of an ultralight scalar field to electromagnetism. The expectation value of the field varies between recombination and today, usually due to the rolling of the scalar, causing a net polarization rotation of light from the CMB. However, this effect is typically avoided in models which predict a small field value in the early Universe. In this work, we demonstrate a robust mechanism to create the birefringence signal regardless of initial conditions. Axions are sourced by the cosmological abundance of nucleons through a CP-violating coupling, which results in the growth of the field if the expectation value at early times is small, and leaves the evolution unaffected if it is large. Measurements of the birefringence angle $\beta$ thus place limits on the combined strength of the photon and nucleon interactions. A slew of upcoming CMB experiments (space-based \cite{LiteBIRD,PICO}, ground-based \cite{Polarbear:2020lii,ACT:2020frw,SPT-3G:2021eoc,BICEP:2021xfz,CLASS,QUBIC:2020kvy,SimonsObservatory:2018koc,Moncelsi:2020ppj,CMBS4_Abazajian}, and balloon-borne \cite{SPIDER:2021ncy,LSPE:2020uos}) will probe this signal with improved calibration accuracy and control over systematic effects. Independent confirmation of recent reports of non-zero $\beta$ would serve as a clear indication of physics beyond the SM, motivating a thorough exploration of models similar to the one discussed in this work.

The monopole-dipole couplings responsible for cosmic birefringence can be generated through renormalizable interactions involving a heavy charged fermion. Inspired by Ref.~\cite{Hook_2018}, we argue that the fine-tuning of the axion mass in such a theory can be avoided by model-building. A compelling alternative would be to source an ultralight scalar through other abundant particles in the Universe rather than SM nucleons. For instance, if neutrinos are Dirac fermions with a scalar Yukawa interaction \cite{Esteban:2022tzx,Pasquini:2015fjv,Xu:2020qek,Babu:2019iml}, a much weaker coupling will suffice to create an observable birefringence signal due to the substantially greater number density of the cosmic neutrino background relative to nucleons. This would result in significantly less fine-tuning, with the interaction expected to be technically natural for cutoffs $\lesssim \text{GeV}$. A detailed evaluation of this possibility, and of similar weakly-constrained couplings to dark matter, would be an interesting subject for future study. 

\acknowledgments
We thank Michael A. Fedderke, David E.~Kaplan, and Surjeet Rajendran for useful discussions.
 
\bibliographystyle{JHEP.bst}
\bibliography{references.bib}

\end{document}